\documentclass[12pt,a4wide]{article}
\usepackage{amsmath,amsthm,epsfig,graphicx}
\usepackage{amsmath,amsthm,amssymb}
\usepackage{amsfonts}
\usepackage{graphicx}
\usepackage{latexsym}
\usepackage[mathcal]{eucal}
\usepackage{epsfig}
\usepackage{verbatim,url}
\newcommand{\Ex}{ \mathbb{E} }

\newtheorem{theorem}{Theorem}[section]
\newtheorem{corollary}{Theorem}[section]

\newtheorem{remark}[theorem]{Remark}

\numberwithin{equation}{section}

\title{\vspace{-3cm}
{\bf {\small An updated version of this paper will appear in the Journal of Financial Engineering, Volume 1, issue 2, 2014, World Scientific. }} \\  \hspace{1cm} \\ {\bf \Large Optimal execution comparison across risks and dynamics, with solutions for displaced diffusions}\thanks{We are grateful to Alexander Schied for helpful comments. This work expresses the opinion of its authors and may not represent the views of the institutions the authors work for}}
\author{Damiano Brigo\\ Dept. of Mathematics\\ Imperial College London\\ and Capco Research Institute \\ {\normalsize {\tt damiano.brigo@imperial.ac.uk}}  \and   Giuseppe Di Graziano \\ Dept. of Mathematics \\ Imperial College London \\ and Deutsche Bank AG, London\\ {\normalsize {\tt giuseppe.di-graziano@db.com}}}

\date{\small{First Version: April 10, 2013. This version: \today}}

\begin{document}

\maketitle

\thispagestyle{empty}

\begin{abstract}
We solve a version of the optimal trade execution problem when the mid asset price follows a displaced diffusion. Optimal strategies in the adapted class under various risk criteria, namely value-at-risk, expected shortfall and a new criterion called ``squared asset expectation" (SAE), related to a version of the cost variance measure, are derived and compared. It is well known that displaced diffusions (DD) exhibit dynamics which are in-between arithmetic Brownian motions (ABM) and  geometric Brownian motions (GBM) depending of the choice of the shift parameter. Furthermore,  DD allows for changes in the support of the mid asset price distribution, allowing one to include a minimum permitted value for the mid price, either positive or negative.  We study the dependence of the optimal solution on the choice of the risk aversion criterion. Optimal solutions across criteria and asset dynamics are comparable although differences are not negligible for high levels of risk aversion and low market impact assets. This is illustrated with numerical examples. 
\end{abstract}

{\bf AMS Classification Codes}: 60H10, 60J60, 91B70;


\medskip



\pagestyle{myheadings}
\markboth{}{{\footnotesize  D. Brigo, G. Di Graziano, Optimal  Execution under Displaced Diffusion and Comparisons}}

\section{Introduction}
Optimal trade execution is an area of financial mathematics that is generating an increasing interest in the literature as well as in the industry.


A fundamental problem algorithmic traders face is the division of a large order into smaller trades to minimise its price impact. Almgren and Chriss formalized and solved the execution problem in a number of seminal papers, notably \cite{almgren1999} and \cite{almgren2000}, by minimizing the expected cost of execution and a cost-variance risk criterion. The authors above assume that the mid price of the asset is modelled by an arithmetic Brownian motion (ABM) and the temporary price impact of an order is proportional to the quantity executed. They find the optimal strategy in the set of deterministic strategies. The advantage of such approach is that it is possible to derive deterministic,  closed-form solutions to the optimal execution
problem. A well known disadvantage of ABM based models is that they allow for negative asset values. 


Gatheral and Schied \cite{gatheral2011} solve a similar problem under the assumption that the mid asset price follows a geometric Brownian motion (GBM), which guaranties positive asset prices, and where the trading strategies are allowed to be random and adapted in particular. However, the combination of GBM dynamics and cost variance risk criterion does not allow for closed form solutions in this setting. 
Numerical studies carried out by Forsyth \cite{forsyth2009} and Forsyth et al \cite{forsyth2009b}, show that the optimal trading rate is almost the same under ABM and GBM dynamics.

By introducing an alternative risk criterion, a version of the value at risk (VaR) measure, Gatheral and Schied \cite{gatheral2011} manage to derive closed form formulae for the optimal execution rate under adapted trading strategies. The authors compare the behaviour of the optimal solution using the value risk criterion when asset prices follow both GBM and ABM price dynamics.

In this paper we model asset prices using a displaced diffusion (DD) process. Displaced diffusions  can in fact give rise to dynamics which are an hybrid between ABM and  GBM depending on the choice of the shift parameter values. In particular, such an approach allows to model volatility as a combination of constant absolute volatility (as in the ABM) and level-proportional absolute volatility (as in GBM). Furthermore, DD allows to set a minimum possible value, either positive or negative, for the mid asset price distribution, which is otherwise zero in GBM or minus infinity in ABM. We look at the trade execution problem for the class of absolutely continuous and adapted trading strategies. 

In this paper we also investigate the impact of different risk criteria on the optimal solution and derive a simple mapping between those criteria. 
We start from a reasonable benchmark case, finding very similar solutions across models and criteria, all departing from the linear Volume Weighted Average Price (VWAP) solution.
Subsequently, we reduce the impact parameter of a factor 10 and find that, for relatively illiquid securities, the optimal execution policy is similar for all models and close to linear.
We also examine the effect of applying a large shift in the DD model, finding that the DD+VaR solution is slower than the equivalent GBM+VaR solution. Finally, we halve the impact parameter wrt the benchmark case and keep a large shift, finding that the VaR based solutions may not be monotone, and in particular may go negative locally. 

Overall we find that optimal solutions across criteria and asset dynamics are comparable, although the differences are not negligible for high levels of risk aversion and low market impact assets.
%

The paper is organized as follows: Section \ref{sec:generaloptex} introduces the optimal execution problem. The  value at risk (VaR) and expected shortfall (ES) risk measures are discussed in detail, and possible choices for the mid asset price dynamics are considered. In particular, the DD choice is analyzed in detail. 
In section \ref{sec:optimaldd} we solve the optimal execution problem when the mid asset price follows a displaced diffusion and the risk function is either the value at risk or the expected shortfall. 

Section \ref{sec:SAE} introduces a simple risk criterion, the squared asset expectation (SAE), which will be later used to facilitate the comparison of the optimal solution under different asset dynamics.
Approximate solutions to the execution problem under the SAE criterion are provided. 
Section \ref{sec:comparing} compares the optimal execution solutions under different criteria and dynamics. A few numerical examples are presented and discussed. 

\section{The trade execution problem under different dynamics}\label{sec:generaloptex}

Assume that an agent, e.g. a broker or a dealer, wishes to buy or sell a given number of asset units (e.g. shares) within a pre-specified time horizon. If the total amount of contracts to execute is relatively high compared to the asset's traded volumes, the transaction is likely to affect its price. The price impact is usually classified as instantaneous, transient or permanent. Optimal strategies aim to find an execution schedule which minimizes such an impact and thus the expected cost of the transaction. Often risk criteria are introduced to take into account the impact of exogenous factors on the price of the asset. For example, the execution of a large sale of a stock may also be adversely affected by negative news regarding the company's profits. In this paper, we shall restrict our attention to continuous execution strategies.  

Let $x(t)$ be the units of the asset yet to be executed at time $t$, with $x(0)=X$ the initial sale size and $x(T)=0$ at the final time $T$, and assume that the price at which the transaction is executed at time $t \in (0,T]$ is given by
\[ \widetilde{S}_t = S_t + \eta\ \dot{x}(t) + \gamma (x(t) - x_0) \]
where $\eta$ and $\gamma$ are constants and $S$ is the unaffected asset price level. 
In line with \cite{gatheral2011}, we assume $x(t)$ trajectories to be absolutely continuous and $x(t)$ to be adapted to the filtration generated by $S$. 
We can identify the following impact terms:

\begin{itemize}
\item $\eta\ \dot{x}(t)$ is the instantaneous impact of trading $d x(t) = \dot{x}(t) dt$ shares in $[t,t+dt)$ and only affects the $[t,t+dt)$ order. 
\item $\gamma (x(t) - x_0)$ represents a permanent component of the price impact that has been accumulating over $[0, t]$ by all transactions up to $t$. 
\end{itemize}


\subsection{ABM vs GBM price dynamics}

Almgren and Chriss \cite{almgren1999} and \cite{almgren2000} assume that the unaffected asset price process $S$ follows an ABM:

\[ d S_t = \sigma S_0 dW_t, \ \ S_0  \ \ \ \mbox{(ABM)}\]
where $W$ is a standard Brownian motion. 

Gatheral and Schied \cite{gatheral2011} on the other hand postulate instead that

\[ d S_t = \sigma S_t dW_t, \ \ S_0  \ \ \ \mbox{(GBM)}. \]

Note that the absolute volatility in the ABM model is proportional to the initial stock price, i.e. $S_0$ whereas in the GBM case, it is proportional to the current price $S_t$.   

When long execution time horizons are considered, the ABM model has the drawback that the probability of the asset price going negative is non negligible. However, execution of trades usually takes place over short time horizons, so the issue is not that severe in real life applications.  Of course, GBM based models do not suffer from this drawback. 

\subsection{Displaced diffusions and price dynamics}

We shall now assume that the unaffected asset price $S$ follows a displaced diffusion.  Let  $Y_t$ be a GBM with volatility $\sigma$ and initial level $Y_0 =S_0 - K$,
\[  d Y_t = \sigma Y_t dW_t \]
Define the asset price $S$ as the sum of $Y_t$ and the constant shift parameter $K$,
\[ S_t = K + Y_t\ \  \mbox{or equivalently} \ \ d S_t = \sigma (S_t-K) dW_t, \ S_0 \ \mbox{(DD)}\]
In order for asset prices to remain positive, we need to assume that $K \ge 0$. In the opposite case, namely  $K < 0$, we may consider the probability that the asset price will be negative at some future time $t$
\begin{equation}
\label{Snegative}
 \mathbb{P}(S_t<0) = \Phi\left(\frac{- \ln(1-S_0/K) + \sigma^2 t /2}{\sigma \sqrt{t}} \right).
 \end{equation}
It is straightforward to see that (\ref{Snegative}) tends to one as $t$ becomes very large. However, one should again keep in mind that typical execution horizons are quite small, so that large times are never seen in practice.

The key advantage of displaced diffusions is that they behave like a combination of arithmetic Brownian motions and geometric Brownian motions. Consider in fact the absolute volatility of the asset in the DD model: 
\begin{equation}
\label{volS}
 \sigma (S_t-K)  = \sigma S_t (1-K/S_t). 
\end{equation}
For large values of $S_t$, the ratio $K/S_t$ becomes small and we have that
\[  \sigma S_t (1-K/S_t) \approx \sigma S_t.\]
which resembles the level-proportional absolute volatility of the GBM.
On the other hand, for small values of $S_t$, expression (\ref{volS}) is closer to the constant absolute volatility of the ABM
\[ \sigma (S_t-K)  \approx \sigma (-K) .\]

One more obvious advantage of DD with respect to ABM and GBM is that we can move the initial point in the support of the asset probability distribution to a chosen point in the positive or negative range. With ABM there is no such choice, as the support begins at $-\infty$, and similarly for GBM, where the support begins at~$0$.

Having introduced the basic dynamics for asset price process, we turn our attention to the optimal execution problem. Following the pioneering approach of \cite{almgren2000}, we formulate the optimal execution problem as a trade off between costs of executing large quantities of the asset in a short period of time and the risk that the mid price moves away if execution is delayed.

\subsection{Cost Function}
The instantaneous cost of executing $d x(t) = \dot{x}(t)dt$ units over the interval $[t,t+dt)$ at price price $\widetilde{S}_t$ is given by $\widetilde{S}_t \dot{x}(t) dt$. Straightforward calculations show that the total execution cost is equal to

\begin{eqnarray*} C(x) :=  \int_0^T \widetilde{S}_t  \dot{x}(t) dt =    \int_0^T [S_t + \eta\ \dot{x}(t) + \gamma (x(t) - x_0)]\dot{x}(t)dt \\
= -X S_0 - \int_0^T x(t) dS_t + \eta \int_0^T (\dot{x}(t))^2 dt + \gamma X^2/2, 
\end{eqnarray*}
where we have made use of the integration by parts formula. The calculation above is valid for any continuous martingale $S$.

\subsection{Risk Function: Value at Risk} 
Various risk functions have been proposed in the literature.  
Gatheral and Schied \cite{gatheral2011} suggest to use a risk measure based on the popular concept of Value at Risk (VaR). In order to see how the concept of VaR can be applied to optimal execution, let $\nu_{\alpha,t,h}$ be the VaR measure at time $t$ for a given confidence level $\alpha$ (e.g. $\alpha = 0.95$)  and time horizon $h$.  The VaR of an outstanding position $x(t)$ at time $t$ over the interval $[t, t+h)$ is given by
\begin{eqnarray*}
\nu_t[ x(t) ( S_t - S_{t+h}) ]  &=& x(t)  \nu_t[ ( S_t - S_{t+h}) ]\\ 
&=& x(t)  \nu_t [ ( Y_t + K - Y_{t+h} - K) ] \\
&=& x(t)  \nu_t  [ ( Y_t  - Y_{t+h}) ]\\
&=& x(t) \nu_t[Y_t(1- \exp(-\sigma^2 h/2 + \sigma (W_{t+h} - W_t)))] \\
&= &x(t) Y_t \nu_t [1- \exp(-\sigma^2 h/2 + \sigma \sqrt{h} \epsilon)]\\
&=& x(t) Y_t q_\alpha[1- \exp(-\sigma^2 h/2 + \sigma \sqrt{h} \epsilon)]
\end{eqnarray*}
where $\epsilon$ is a standard normal random variable and  $q_\alpha(X)$ is the $\alpha$ quantile of the distribution  of the random variable $X$. Note that we have used the homogeneity property of the VaR. 

Furthermore, define the constant
\begin{eqnarray*}
\tilde{\lambda}_\alpha := q_\alpha[1- \exp(-\sigma^2 h/2 + \sigma \sqrt{h} \epsilon)] = 1- \exp(-\sigma^2 h/2 + \sigma \sqrt{h} q_{1-\alpha}(\epsilon)),
\end{eqnarray*}
where $q_{1-\alpha}(\epsilon)$ can be inferred from the normal distribution tables, 
to obtain that
\begin{eqnarray*}
 \nu_t[ x(t) ( S_t - S_{t+h}) ]  = \tilde{\lambda} x(t) Y_t = \tilde{\lambda} x(t) (S_t - K).
\end{eqnarray*}
From the calculations above we can derive the VaR over the life of the strategy by integrating over the interval $[0,T]$ to obtain 
\begin{equation}
\label{RVaR}
R^{\mbox{\footnotesize VaR}_\alpha}(x) := \tilde{\lambda} \int_0^T x(t) (S_t - K) dt.
\end{equation} 
Equation (\ref{RVaR}) is the equivalent for displaced diffusions of the average VaR measure derived by \cite{gatheral2011}.

\subsection{Risk Function: Expected Shortfall}
An alternative interesting risk measure is the expected shortfall. In particular the expected shortfall $\mu_{\alpha,t,h}$ at time $t$ for the horizon $t+h$  and confidence level $\alpha$ is defined as the expected value of the loss of the position beyond the VaR level, i.e.
\[ \mu_{\alpha,t,h} := \mathbb{E} \{ S_t - S_{t+h} | S_t - S_{t+h} \ge \nu_{\alpha,t,h}, {\cal F}_t  \}  \]
where ${\cal F}_t$ is the market filtration at time $t$.
Expected shortfall is sometimes preferred to VaR because it is a coherent risk measure, namely it is sub-additive, while Value at Risk is not. However, recent literature on elicitability shows that Value at Risk is elicitable whereas expected shortfall is not, which is a clear advantage of VaR.  

By proceeding similarly to the VaR case, we have that the expected shortfall over the interval $[t,t+h)$ is equal to 
\begin{eqnarray*}
\mu_t[ x(t) ( S_t - S_{t+h}) ]  
= x(t) Y_t\  m_\alpha[1- \exp(-\sigma^2 h/2 + \sigma \sqrt{h} \epsilon)]
\end{eqnarray*}
where $\epsilon$ is a standard normal random variable and we have used the homogeneity property of the expected shortfall. Straightforward calculations lead to the following result,
\[ m_\alpha = \frac{1}{1-\alpha}\left[ \Phi\left( \frac{\ln(1-\tilde{\lambda}_\alpha) + \sigma^2 h / 2  }{\sigma \sqrt{h}} \right) -  \Phi\left( \frac{\ln(1-\tilde{\lambda}_\alpha) - \sigma^2 h / 2  }{\sigma \sqrt{h}} \right)\right] =: \hat{\lambda}_\alpha \]
and hence
\begin{eqnarray*}
\mu_t[ x(t) ( S_t - S_{t+h}) ]  
= x(t) Y_t\  \hat{\lambda}_\alpha = \hat{\lambda}_\alpha x(t) (S_t - K) .  
\end{eqnarray*}
Averaging over the life of the strategy one obtains
\[  R^{\mbox{\footnotesize ES}_\alpha}(x) := \hat{\lambda} \int_0^T x(t) (S_t - K) dt . \]

\subsection{Cost plus Risk}
We now have all the ingredients to define the optimization problem. Adding up the expected execution cost and the risk function, we obtain
\begin{equation}
\label{costPlusRisk}
\mathbb{E}\left[ C(x) + R(x) \right] = \mathbb{E}\left[ C(x) + L \bar{\lambda} \int_0^T x(t) (S_t - K) dt \right] . 
\end{equation}
In this criterion:
\begin{itemize}
\item $L$ is a cost/risk leverage parameter that measures risk aversion in executing the order. The larger $L$, the more important the risk component of the criterion compared to the cost component; 
\item $\bar{\lambda}$ is either $\tilde{\lambda}$ or $\hat{\lambda}$ depending on whether we are using the VaR or expected shortfall risk function. 
\end{itemize}

Substituting the explicit expression for $C(x)$ derived in the previous section into (\ref{costPlusRisk}), we obtain 
$$ 
 \mathbb{E}\left[ C(x) + R(x) \right] = - X S_0 + \gamma X^2 /2 + \eta \mathbb{E}\left[\int_0^T \dot{x}(t)^2 dt  + L \check{\lambda}  \int_0^T x(t) Y_t dt  \right]
 $$
The optimal execution problem is thus equivalent to the following minimization
\begin{equation}\label{eq:optimumexdd} x^\ast = \mbox{arginf}_{x} \mathbb{E}\left[\int_0^T \dot{x}(t)^2 dt  + L \check{\lambda}  \int_0^T x(t) Y_t dt  \right],
\end{equation}
where we have set $\check{\lambda} =  \bar{\lambda}/\eta$.

\section{Solution of the trade execution problem for displaced diffusions}\label{sec:optimaldd}

Problem (\ref{eq:optimumexdd}) has been solved by Gatheral and Schied \cite{gatheral2011} when the mid price of the asset follows a GBM.  The authors above derive the following explicit solution for the optimal trading strategy   
\[ x^\ast(t) = \frac{T-t}{T} \left[ X -  L \check{\lambda} \frac{T}{4} \int_0^t Y_u du \right] . \]

The risk adjusted cost function can be also calculated explicitly for the optimal strategy and it is equal to
\[\mathbb{E}\left[\int_0^T \dot{x}^\ast(t)^2 dt  + L \check{\lambda}  \int_0^T x^\ast(t) Y_t dt  \right] = \frac{X^2}{T} + \frac{ \check{\lambda} L T X Y_0}{2} - \frac{  (L \check{\lambda})^2}{8 \sigma^6}Y_0^2 \left( e^{\sigma^2 T} - 1 - \sigma^2 T - \frac{\sigma^4 T^2 }{2}\right) \]

Similar calculations in the case of displaced diffusions lead to the following result
\begin{corollary} {(\bf Optimal execution strategy for a displaced diffusion).}
The unique optimal trade execution strategy attaining the infimum in (\ref{eq:optimumexdd}) is
\[ x^\ast_t = \frac{T-t}{T} \left[ X -  \check{\lambda} L \frac{T}{4} \int_0^t (S_u - K) du \right]  \]
Furthermore, the value function for the optimal strategy is given by
\[\mathbb{E}\left[\int_0^T \dot{x}^\ast(t)^2 dt  +  \check{\lambda} L \int_0^T x^\ast(t) (S_t - K) dt  \right] \]\[= \frac{X^2}{T} + \frac{L \check{\lambda} T X (S_0 - K)}{2} - \frac{ (L\check{\lambda})^2}{8 \sigma^6}(S_0-K)^2 \left( e^{\sigma^2 T} - 1 - \sigma^2 T - \frac{\sigma^4 T^2 }{2}\right) \]
\end{corollary}

%

%
%
%
%
%

\section{Optimal solution for displaced diffusion processes under an alternative square risk function}\label{sec:SAE}
In order to test the robustness of the optimal strategy, we consider the alternative "squared-asset expectation" (SAE) risk criteria

\begin{equation}
\label{riskCriteria2}
R^{SAE}(x) \equiv  \lambda \int_0^T  x^2(t) \sigma^2 E[S_t^2] dt.
\end{equation}

\begin{remark} It is important to notice that while in the case of VaR and expected shortfall $\lambda$ can be calculated endogenously, here $\lambda$ is an exogenous parameter.
\end{remark} 

The criterion above is similar to the one used by Almgren and Chriss in \cite{almgren2000}, while differring in an important way. In (\ref{riskCriteria2}) we have replaced the square of $S_t$ with its expectation. Furthermore, we will consider cases where $S_t$ is a displaced diffusion instead of an ABM. Note that the term 
$$g(t) \equiv E[S_t^2]$$ 
is increasing in $t$, whereas the corresponding term under ABM is constant and equal to $S^2_0$.  Note also that in our earlier specification for the ABM framework, the volatility of the asset $\tilde \sigma$ is expressed as a percentage number, since it contains $S_0$, rather than as absolute units of the reference currency. 
Finally, note that the expected cost is the same whether $S_t$ is modelled by a DD or an ABM, so the optimal solution depends on the chosen dynamics only via the choice of the risk function.

The optimal execution strategy can be derived by solving the optimization problem
\begin{equation}
\label{optCriteriaDDAC} x^\ast = \mbox{arginf}_{x} \mathbb{E}\left[\int_0^T \dot{x}(t)^2 dt  + L \lambda  \int_0^T \sigma^2 x^2(t) g(t) dt  \right]. 
\end{equation}
Using calculus of variations, we show in appendix \ref{calcVarProof} that the optimal solution needs to satisfy the ODE
\begin{equation}
\label{ODE}
\ddot{x}(t)=k^2 g(t) x(t),
\end{equation}
where
$$
k \equiv \sigma \sqrt{L \lambda},
$$
and the initial and terminal conditions are given by $x(0)=X$ and $x(T)=0$ respectively. A solution to the boundary value problem above could be found using standard numerical routines.  However, in appendix \ref{numsol} we provide a simple and efficient alternative algorithm based on series expansions.

\begin{remark} {\bf SAE makes the optimal adapted strategy $x(t)$ deterministic.} 
\end{remark}

\section{Comparing Risk Criteria}\label{sec:comparing}
In this section we derive a couple of simple techniques to map the $\lambda$'s of different criteria in such a way that the relative weight of the risk function is of the same order of magnitude.
%

\begin{itemize}
\item A first possible approach is to match the values of the two different risk criteria by assuming that $x(t)=x(0)$ throughout the whole execution time horizon, 
\[ \Ex[ R^{SAE}(x_0)  ] =   \Ex[ R^{\mbox{\tiny{VaR/ES}}_{\alpha}}(x_0)  ]      \]
where $x_0(t) = x(0)$ for all $t\ge 0$, which leads to 
\[  \lambda^{SAE}_{ABM,DD} = \check{\lambda} \  \frac{  S_0 - K}{X \sigma^2 S_0^2}  , 
\]

\item A slightly more refined approach is to assume that $x(t)$ is equal to the VWAP solution, i.e.
\[ x_0(t) = X \frac{T-t}{T} \]
and equalise the corresponding risk functions,
\[ \Ex[ R^{SAE}(x_0)  ] =   \Ex[ R^{\mbox{\tiny{VaR/ES}}_{\alpha}}(x_0)  ]      \]
leading to
\[ \mathbb{E}_0 \left[\int_0^T \check{\lambda} \  x_0(t)\  (S_t - K) dt \right]=   \int_0^T \lambda^{SAE}_{DD} \  x_0^2(t)\  \sigma^2 \mathbb{E}_0[S_t^2] dt  . \]  
Straightforward calculations lead  to following result
\begin{equation}\label{eq:lambdaeqz} \lambda^{SAE}_{DD} = \frac{\check{\lambda}}{2X \sigma^2} \  
\frac{1}{(S_0-K) \left(\frac{\sigma^2 T}{12} + \frac{S_0^2}{3 (S_0-K)^2}  \right) .} 
\end{equation}
where we have used the approximation $e^{\sigma^2 t} \approx 1 + \sigma^2 t$. 
For simplicity we have also set 
\begin{equation}\label{eq:lambdaeqz2} \lambda^{SAE}_{ABM} := \lambda^{SAE}_{DD} . 
\end{equation}

\end{itemize}
%
%

\subsection{Comparative Results and Conclusions}
In this section we present a series of numerical results which compare the different models and risk criteria analyzed in the paper. In the graphs  below,  we compare the optimal solution of the following dynamics/risk criterion combinations:
ABM+SAE, DD+SAE, GBM+VaR and DD+VaR. The parameter $\lambda$ has been calibrated using Equations (\ref{eq:lambdaeqz}) and (\ref{eq:lambdaeqz2}). 

In each example we visualize just one specific path for the asset $S$ and the related optimal strategies, but this is representative of a more comprehensive analysis we ran through many scenarios, so that our conclusions are sufficiently general. 

Consider the sale of $X = 10^6 = 1$ million units of an asset with current price equal to $S_0 = 100$ currency units (e.g. USD). The execution time horizon is one day, i.e. $T=1$ day $=1/252$ years and the volatility of the asset is $ \sigma = \sigma_{1y} = 30\% = 0.3$ annualized or $\sigma_{1d} = 1.89\%$  per day.

\begin{figure}[h!]
\begin{center}
\caption{Model Comparison: $S_0 =100$, $X=10^6$, $T=1$ day, $K=5$, $\sigma_{1d} = 0.0189$, $\eta =20^{-6}$, $L=100$ . The shift parameter $K$ is relatively low. GBM+VaR and DD+VaR give rise to similar solutions, both departing from the linear case.}
\includegraphics[scale=0.85]{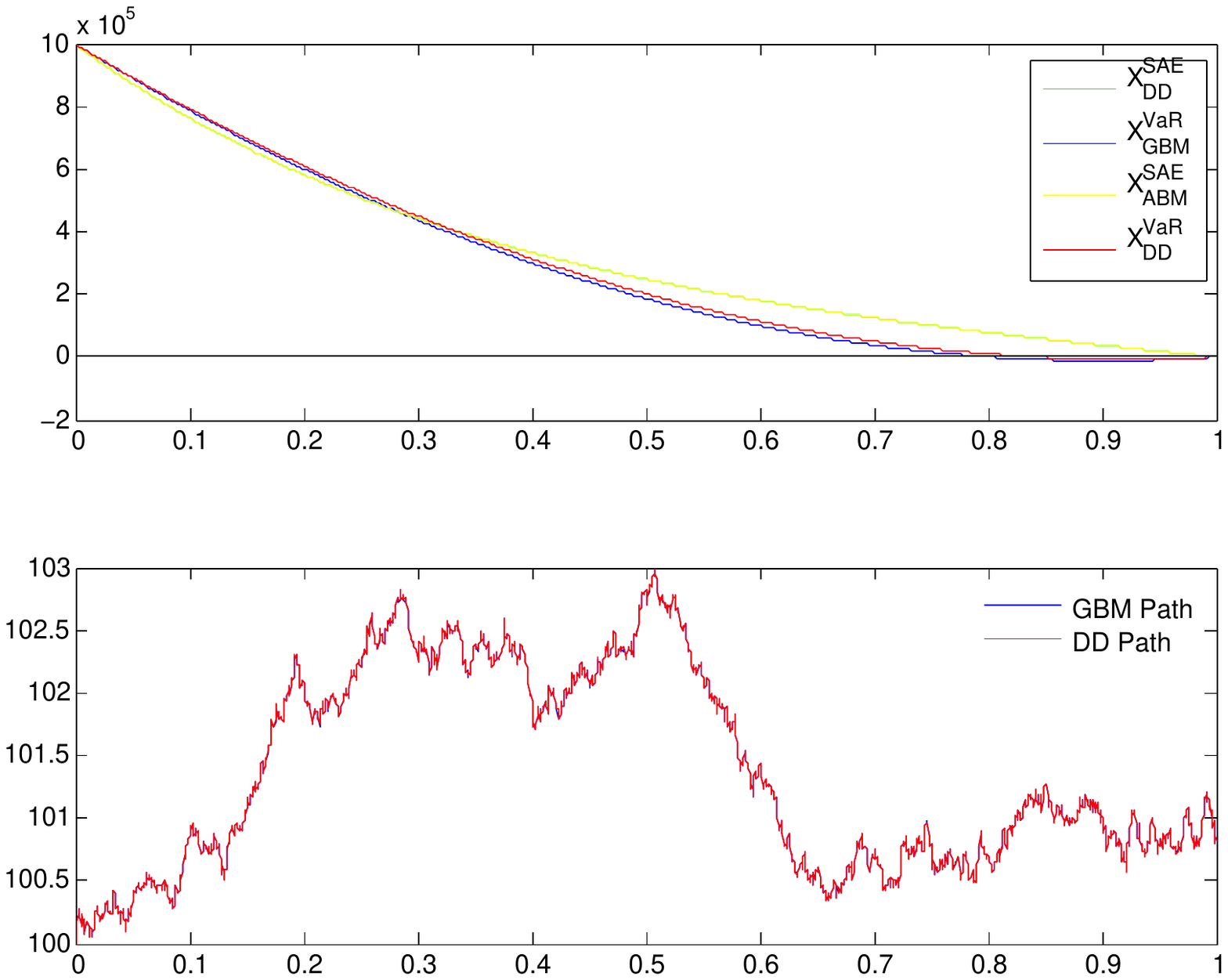}
\label{fig1}
\end{center}
\end{figure}

In Figure \ref{fig1} we show the problem solution when we set the DD shift parameter to $K=5$. The market impact parameter is $\eta=20^{-6}$,  i.e. the price impact of an instantaneous sale of $1$ million units of the asset would be equal to $2$ dollars per asset, or $2\%$ of the mid price.  The cost/risk parameter $L$ is equal to $100$.  For the above choice of parameters, the optimal execution is faster for the model that employs the VaR risk function compared to the SAE based model.

\begin{figure}[h]
\begin{center}
\caption{Model Comparison after reducing the impact parameter $\eta$ of a factor 10 wrt Fig \ref{fig1}: $S_0 =100$, $X=10^6$, $T=1$ day, $K=5$, $\sigma_{1d} = 0.0189$, $\eta =20^{-5}$, $L=100$. For relatively illiquid securities, i.e. when the market impact parameter is high, the optimal execution policy is similar for all models and close to linear. }
\includegraphics[scale=0.85]{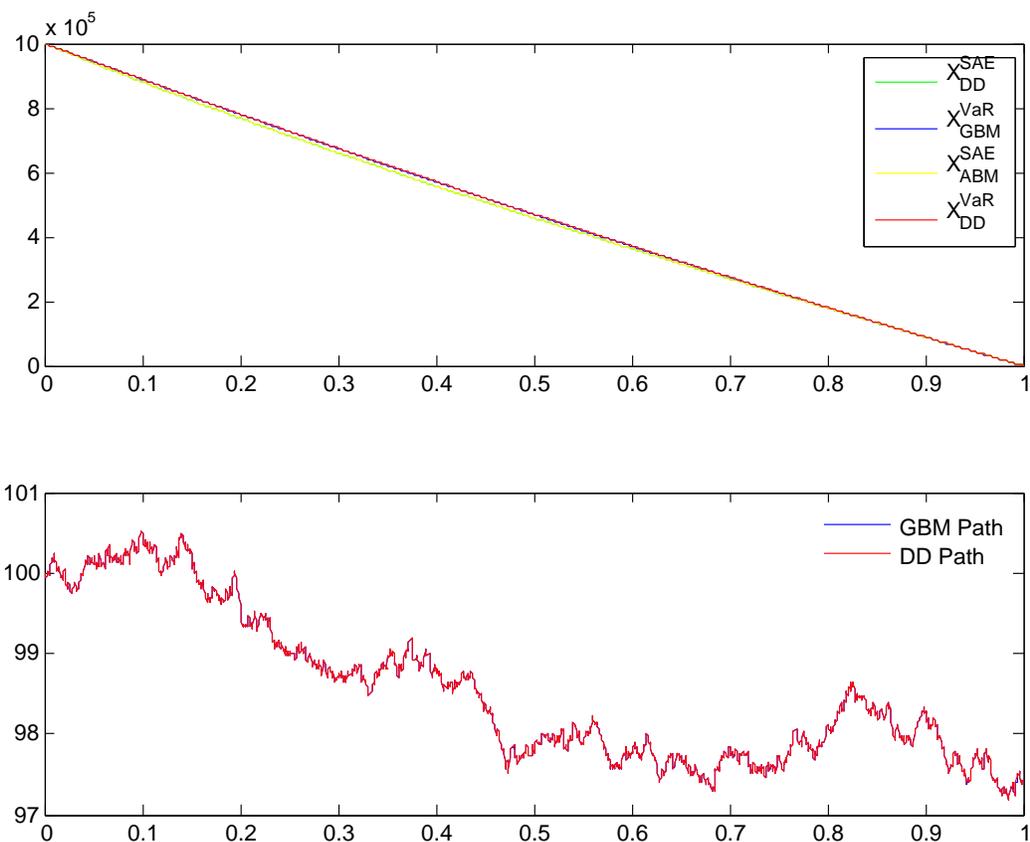}
\label{fig2}
\end{center}
\end{figure}

Figure \ref{fig2} is based on the same parameters as in Figure \ref{fig1} with the exception of the impact parameter, which is increased by a factor of $10$ to $\eta = 20^{-5}$. This is an example of a relatively illiquid security. In this case, the price impact dominates over the risk component and the execution strategy is very close to linear for all models. 

\begin{figure}[h]
\begin{center}
\caption{Model Comparison after increasing the shift $K$ to $50$ wrt Fig \ref{fig1} : $S_0 =100$, $X=10^6$, $T=1$ day, $K=50$, $\sigma_{1d} = 0.0189$, $\eta =20^{-6}$, $L=100$. For high levels of the shift parameter $K$, the DD+VaR solution is slower than the equivalent GBM+VaR solution }
\includegraphics[scale=0.85]{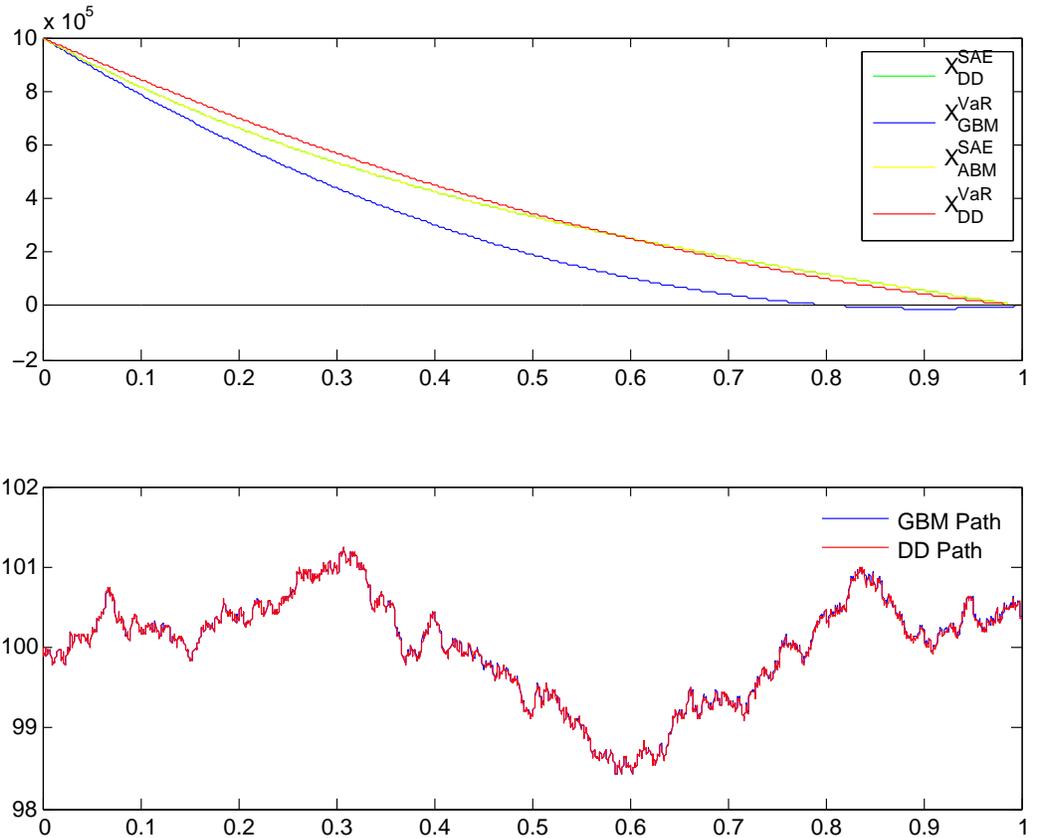}
\label{fig3}
\end{center}
\end{figure}

In Figure \ref{fig3} we used the same parameters as in Figure \ref{fig1} but increased the DD shift parameter to $K=50$. The optimal execution policy is faster for the GBM+VaR combination compared to DD+VaR. This is due to the fact that in the first case the risk function depends on the path of $S_t$ whereas in the second case it depends on $S_t-K$, which is roughly half in order of magnitude. However,  note that the instantaneous volatility of the DD is rescaled using Formula (\ref{rescaledVol}) below, to ensure that the integrated volatility of the DD and GBM models are of similar order of magnitude,

\begin{equation}
\label{rescaledVol}
\sigma^{DD} = \sigma^{GBM} \frac{S_0}{S_0-K}.
\end{equation}

\begin{figure}[h]
\begin{center}
\caption{Model Comparison after halving the impact parameter $\eta$ and increasing the shift $K$ to $50$ wrt Fig \ref{fig1}: $S_0 =100$, $X=10^6$, $T=1$ day, $K=50$, $\sigma_{1d}= 0.0189$, $\eta =10^{-6}$, $L=100$. For relatively low levels of $\eta$ and high levels of $L$, the VaR based solutions may not be monotone.}
\includegraphics[scale=0.85]{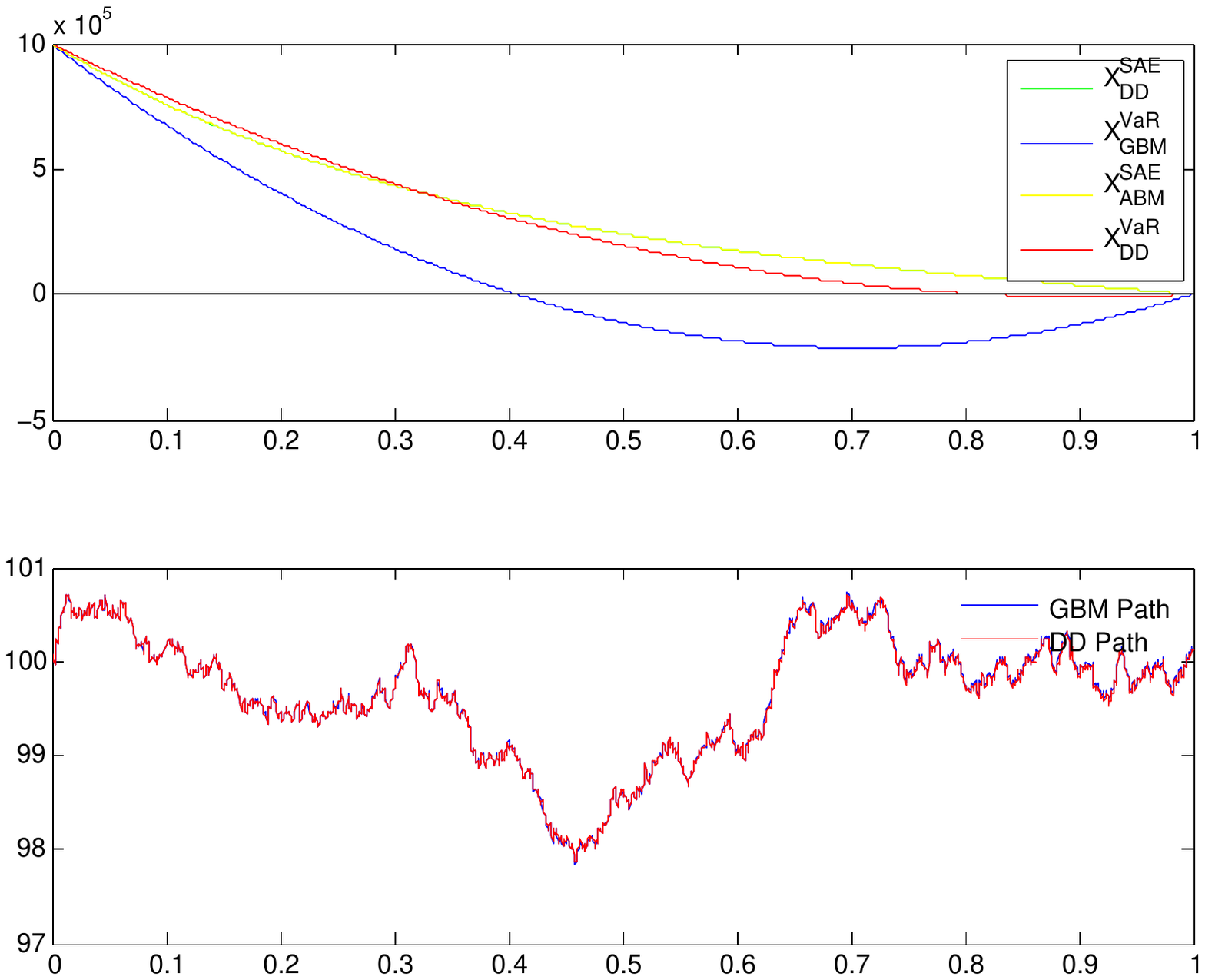}
\label{fig4}
\end{center}
\end{figure}

In Figure \ref{fig4} we reduced the impact parameter $\eta$ to half its previous value, namely $10^{-6}$. Note that in this case the GBM+VaR solution is not monotone as no smooth constraint at zero has been imposed on $x_t$. Note also that a similar behaviour may occur with the  DD+VaR combination. 

From the examples above, it emerges that for high levels of risk aversion and relatively low impact, the optimal execution may be significantly different for the linear solution. Different combinations of asset dynamics and risk function give rise to different solutions in these cases, although this is partly due to the specific choice of parameters mapping between different models and of the subjective parameter $L$. 

%
%

\newpage

\section*{Appendices}

\appendix

\section{Derivation of the ODE for the Optimal Strategy\label{calcVarProof}}

The optimisation problem (\ref{optCriteriaDDAC}) requires that we find the path $x(t)$ which minimises
\begin{equation}
\label{lossFunc}
L[x] \equiv \mathbb{E}\left[\int_0^T \dot{x}(t)^2 dt  + L \lambda  \int_0^T \sigma^2 x^2(t) g(t) dt  \right].
\end{equation}
In order to do so,  consider the following perturbation of the processes $x(t)$ and $\dot x(t)$

\begin{eqnarray*}
x^{\epsilon}(t) &=& x(t) + \epsilon h(t)\\
\dot x^{\epsilon}(t) &=& \dot x(t) + \epsilon \dot h(t)
\end{eqnarray*}
where $h(t)$ is an arbitrary function satisfying $h(0)=h(T)=0$ and $\epsilon$ is a real constant. Substituting the perturbed path into (\ref{lossFunc}) we obtain

$$
H(\epsilon)  = \mathbb{E}\left[ \int_0^T (\dot x^{\epsilon}(t))^2 dt + L \lambda \int_0^T \sigma^2 (x^{\epsilon}(t))^2 g(t) dt\right].
$$
Taking the first derivative with respect to $\epsilon$ and evaluating the resulting expression at $\epsilon=0$, it follows that
\begin{equation}
\label{Hprime}
\dot H(0)  = \mathbb{E}\left[2 \int_0^T \dot x(t) \dot h(t) dt + 2 L \lambda \int_0^T \sigma^2 x(t)h(t) g(t) dt\right].
\end{equation}
Using the integration by parts formula and the constraint $h(0)=h(T)=0$, we obtain the equality
$$
\int_0^T \dot x(t) \dot h(t) dt = -\int_0^T \ddot x(t)h(t) dt,
$$
which allows us to simplify (\ref{Hprime}) as follows 
$$
\dot H(0)  = \mathbb{E}\left[2 \int_0^T\left( \ddot x(t)  - L \lambda  \sigma^2 x(t)g(t)\right)h(t)  dt\right].
$$

The optimal path $x(t)$ is obtained by setting $\dot H(0)=0$. Since the function $h(t)$ was chosen arbitrarily, the following must hold for all $0\leq t \leq T$ 
 $$
 \ddot x(t)  - L \lambda  \sigma^2 g(t) x(t)=0.
$$
This ODE is deterministic even if our setting allows for an adapted $x(t)$. 

\section{Numerical algorithm for optimal solutions\label{numsol}}
Consider the ODE
$$
\ddot{x}(t)=k^2 g(t) x(t).
$$
Simple manipulations show that $g(t)$ can be written as follows

\begin{eqnarray}
g(t) & =& Y_0^2 \left[ \left(\frac{K}{Y_0}\right)^2+2\left(\frac{K}{Y_0}\right)+\exp(\sigma^2 t)\right]\\
         &=& \bar k^2 (\alpha_0 +e^{\sigma^2 t})
\end{eqnarray}
where 
$$\bar k^2 \equiv (Y_0 k)^2$$ 
and 
$$
\alpha_0 \equiv \left(\frac{K}{Y_0}\right)^2+2\frac{K}{Y_0}.
$$

Assume that the solution of the ODE can be expanded in a Taylor series around $t=0$

\begin{equation}
\label{TaylorSolution}
x(t) = \sum_{j=0}^{\infty} a_j t^j.
\end{equation}
The coefficient of the series can be found by substituting (\ref{TaylorSolution}) into the ODE,

$$
\sum_{j=0}^{\infty} (j+2) (j+1) a_{j+2} t^j = \bar k^2 \left(\alpha_0+ \exp(v t)\right) \left(\sum_{j=0}^{\infty} a_j t^j\right).
$$

Substituting the exponential term with its Taylor expansion in $t$ and exchanging the order of summation we obtain
 
\begin{eqnarray}
\sum_0^{\infty} (j+2) (j+1) a_{j+2} t^j &=& {\bar k}^2 (1+\alpha_0) \left(\sum_{j=0}^{\infty} a_j t^j\right)+ {\bar k}^2  \sum_{i=1}^{\infty}  \sum_{j=0}^{\infty} \frac{v^i}{i!} a_j t^{j+i} \nonumber \\
&=&  {\bar k}^2 (1+\alpha_0) \left(\sum_{j=0}^{\infty} a_j t^j\right)+ {\bar k}^2  \sum_{i=1}^{\infty}  \sum_{j=i}^{\infty}  \frac{v^i}{i!} a_{j-i} t^{j}. 
\end{eqnarray}
 
Truncating both sums at $m$ and equating the coefficient of the same order in $t$ on both sides of the equation we obtain a recursive expression for the coefficients $a_j$,
 
$$
a_{j+2} = \frac{\bar k^2}{(j+2)(j+1)} \left( (1+\alpha_0) a_j + \sum_{k=1}^j a_{j-k} \frac{v^k}{k!} \right). 
$$
for $j=0$ to $m-2$.

The expression above can be written in a more compact form using vector notation

$$
a_{j+2} = \frac{\bar k^2}{(j+2)(j+1)} \left( (1+\alpha_0) a_j,  A_{j-1}\right) \cdot V_j
$$

where we have defined
$$
A_j \equiv \left(a_0, a_1, \dots, a_j\right)^{'}
$$

$$
V_j \equiv \left(1, v, \frac{v^2}{2!} \dots, \frac{v^j}{j!}\right)^{'} 
$$

Using the initial condition $x(0)=X$ it follows that $a_0 = X$. On the other hand, $a_1$ can be derived numerically via a simple root search using the terminal condition $x(T) =0$.


\begin{thebibliography}{99}


\bibitem{alfonsi2010}  Alfonsi, A. and A. Schied (2010). Optimal Trade Execution and Absence of Price Manipulations in Limit Order Book Models.
SIAM Journal on Financial Mathematics, Vol. 1, No. 1, pp. 490-522

\bibitem{alfonsi2012} Alfonsi, A., Schied, A., and A. Slynko (2012).
Order Book Resilience, Price Manipulation, and the Positive Portfolio Problem.
SIAM Journal on Financial Mathematics, Vol. 3, No. 1, pp. 511-533

\bibitem{almgren1999} Almgren, R., Chriss, N. (1999). Value under liquidation. Risk, Dec. 1999.

\bibitem{almgren2000}
 Almgren, R., Chriss, N. (2000). Optimal execution of portfolio transactions. J. Risk 3, 5-39 (2000).

\bibitem{almgren2012}
Almgren, R. (2012). Optimal Trading with Stochastic Liquidity and Volatility. 
SIAM Journal on Financial Mathematics, Vol. 3, No. 1, pp. 163-181

\bibitem{brigodigrazianoarxiv}
Brigo, D., and Di Graziano, G. (2013). Optimal execution comparison across risks and
dynamics, with solutions for displaced diffusions. Available at http://arxiv.org/abs/1304.2942 and 
http://ssrn.com/abstract=2247951

\bibitem{gatheral2011} Gatheral, J., and Schied, A. (2011). Optimal Trade Execution under Geometric Brownian Motion in the Almgren and Chriss Framework. International Journal of Theoretical and Applied Finance, Vol. 14, No. 3, pp. 353--368

\bibitem{forsyth2009}
Forsyth, P. (2009). A Hamilton Jacobi Bellman approach to optimal trade execution. Preprint, available at {\tt http://www.cs.uwaterloo.ca/$\sim$paforsyt/optimal trade.pdf}

\bibitem{forsyth2009b}
Forsyth, P., Kennedy J., Tse T.S., Windcliff H. (2009). Optimal Trade Execution: A Mean-Quadratic-Variation Approach. Preprint (2009) available at
{\tt http://www.cs.uwaterloo.ca/$\_$paforsyt/quad trade.pdf}

\bibitem{schiedwp} Schied, A. (2012). 
Robust Strategies for Optimal Order Execution
in the Almgren--Chriss Framework. Working Paper. 

\end{thebibliography}
\end{document}